\documentstyle[12pt]{article}

\topmargin -\headheight

\def\CC{{\bf C}}
\def\Ker{\mathop{\rm Ker}\nolimits}
\def\Im{\mathop{\rm Im}\nolimits}

\begin{document}

\begin{titlepage}
   
\title{Minimal Models of Integrable Lattice Theory\\
and Truncated Functional Equations}
\author{{\sc A.~BELAVIN}\\
{\em Landau Institute for Theoretical Physics}\\
{\em Chernogolovka, Moscow region 142432, Russia}
\vspace{0.5cm}\\
	{\sc Yu.~STROGANOV}\\
{\em Institute for High Energy Physics}\\
{\em Protvino, Moscow region, Russia}
\vspace{0.5cm}\\
{\em Research Institute for Mathematical Sciences\footnote{till April 2000}}\\
{\em Kyoto University, Kyoto 606, Japan}}
\date{hep-th/9908050}

\end{titlepage}

\maketitle

\begin{abstract}
We consider the integrable XXZ model with the special open 
boundary conditions. We perform Quantum Group reduction 
of this model in roots of unity and use it for the definition 
of Minimal Models of Interable lattice theory. It is shown that 
after this Quantum Group reduction  Sklyanin's transfer-matrices 
satisfy the closed system of the truncated functional relations.
We solve these equations for the  simplest case.
\end{abstract}

\setcounter{page}{1}

\vspace{1.3cm}

\section*{I. Introduction}
		
One of the most interesting open problem in Conformal Field Theory
is the connection between its Integrable Structure~\cite{BLZ}  and
the Fermi--Bose correspondence. This correspondence manifests itself
in the generalizations of Roger--Ramanujian Identities for Characters
of Minimal Models of CFT~\cite{BPZ} discovered by Kedem, Klassen, McCoy 
and Melzer~\cite{KKMM}.
The simplest example of such identity is the following equation
for a character of $M(3,4)$ minimal model(Ising model):  
\begin{equation}
   \label{eq:a}
   \frac{1}{\prod_{n=1}^{\infty}(1-t^n)}
   \sum_{k=-\infty}^{\infty}
   [t^{k(12k+1)}-t^{(4k+1)(3k+1)}]
  =\sum_{\scriptstyle m\>-\>{\rm even}\atop
\scriptstyle m>0}^{\infty}\frac{t^{m^2/2}}{(1-t)\ldots(1-t^m)}
\end{equation} 
The l.~h.~s.\ of the identity has a natural explanation
as Feigin--Fuchs-Felder representation of the space of states 
of the model in terms of free bosonic field.
The r.~h.~s.\ can be interpreted~\cite{KKMM} as a partition
function of right moving (chiral) fermions of one kind.

The similar identities were found in~\cite {KKMM} 
for generic Minimal Model $M(p,p+1)$ with $ p \geq 3 $.
In this case the r.~h.~s.\ can be considered as partition functions for
$ p-2 $ species of right moving fermions.
It was  shown by R.Kedem and B.McCoy tahat the fermionic  states arise as
low lying excitations in Bethe Ansatz approach for the 3-Potts model
which corresponds to $M(5,6)$ Minimal Model. It would be
interesting to generalize these results for general $M(p,p+1)$ case.

In~\cite{BLZ} the continuous field theory version of the
commuting transfer-matrices of Integrable lattice theory
have been constructed and the conjecture that eigenvectors
of these transfer-matrices form the fermionic basis of Virasoro 
characters has been claimed.

So the investigation of the connection between Fermi-Bose correspondence
and Integrable Structure of CFT continues to be interesting both 
in the continuous field theory and in the Integrable   lattice theory
and is one of the main motivations of this work~\cite{BF}.

The plan of the paper is the following. In part~II we consider
the integrable XXZ model with special open boundary conditions which 
was investigated firstly by Alcaraz {\it et al.}~\cite{ALC}.
According to
Pasqier and Saleur~\cite{PS} we use the Quantum Group reduction 
of this model for the definition Minimal Models of Interable lattice 
theory ($LM(p,p+1)$) which coincide with Minimal Models of CFT $M(p,p+1)$ 
in the termodinamic limit. In part~III we show that after this Quantum Group
reduction Sklyanin's transfer-matrix $T(u)$ satisfies a closed system 
of the Truncated Functional relations.
In part~IV we solve these equations for the case $LM(3,4)$.

\section*{II. Minimal Models of Integrable Lattice Theory}	 

Let us consider one dimensional XXZ chain with free
boundary conditions~\cite{ALC}
\begin{eqnarray}
   \label{eq:b}
   H_{xxz}&=&\sum_{n=1}^{N-1}[\sigma_n^{+}\sigma_{n+1}^{-}
   +\sigma_n^{-}\sigma_{n+1}^{+}
   +\frac{q+q^{-1}}{4}\,\sigma_n^z\sigma_{n+1}^z\nonumber\\
   & &\qquad
   +\frac{q-q^{-1}}{4}\,(\sigma_n^z-\sigma_{n+1}^z)]\\
   \sigma_n^\pm&=&1\otimes\ldots\otimes\sigma^\pm
   \otimes\ldots\otimes1\nonumber\\
   \sigma_n^z&=&1\otimes\ldots\otimes\sigma^z
   \otimes\ldots\otimes1\nonumber\\
   \sigma^{+}&=&\pmatrix{
      0&1\cr
      0&0},
\quad
   \sigma^{-}=\pmatrix{
      0&0\cr
      1&0},
\quad
   \sigma^z=\pmatrix{
      1&0\cr
      0&-1}
\nonumber
\end{eqnarray}

It was observed in~\cite{ALC,HAM} that the eigenvalues
of $2L$-site XXZ Hamiltonian~(\ref{eq:b}) with  the coupling
$\eta=\pi/4$ and $\eta=\pi/6$ (we denote $q\equiv e^{\imath\eta}$)
 are exactly coincide with some of the eigenenergies of 
$L$-site self dual quantum Ising and 3-Potts models with free ends
(see also~\cite{ABF}).

The remarkable properties of the model are connected with its 
$U_q(sl(2))$ symmetry found by Pasqier and Saleur~\cite{PS}.
Namely Hamiltonian $H_{xxz}$ commutes with the generators  $X$, $Y$, $H$ of
this quantum algebra defined as follows
\begin{eqnarray}
   & &  X=\sum_{n=1}^N q^{\frac{1}{2}(\sigma_1^z+\ldots+\sigma_{n-1}^z)}
 \sigma_n^{+}
q^{-\frac{1}{2}(\sigma_{n+1}^z+\ldots+\sigma_{N}^z)}\nonumber\\
   & &
   X\to Y,\quad \sigma^{+}\to\sigma^{-}\nonumber\\
   & &
   H=\sum_{n=1}^N\frac{\sigma_n^z}{2};\nonumber
\end{eqnarray}
and satisfy the following relations
\begin{equation}
   \label{eq:d}
   [H,X]=X,\quad [H,Y]=-Y,\quad
   [X,Y]=\frac{q^{2H}-q^{-2H}}{q-q^{-1}}
\end{equation}
Due to the Quantum Group symmetry the spectrum of the Hamiltonian
can be classified according to the Represenation Theory of the algebra.
For generic $q$ its representations are equivalent to the ones of 
the ordinary $ U(sl(2)) $ algebra and the configuration space $(\CC^2)^N$
of the spin chain can be splited into a direct sum of irreducible
highest weight representations $ \rho _j $ ($j$ is the highest weight)
which are in one-to-one correspondence to the ordinary $ sl(2)$ 
representations.
For example in $N=4$ case $(\CC^2)^4$ can be decomposed as 
$ \rho_2 + 3\rho_1 + 2\rho_0 $.

We will concentrate on the case $q^{p+1}=-1$~\cite {LUS,PS}.
In this case the generators $X$ and $Y$ are nilpotent
on the space of states of the model
\begin{equation}
 \label{eq:e}
   X^{p+1}=0, \quad Y^{p+1}=0  	 
\end{equation}	  
As the consequence we obtain the very different picture of decomposition
the configuration space $(\CC^2)^N$.

For example if $q^4=-1$ and we try to decompose $(\CC^2)^4$ we get
the following~\cite{PS}.
The $(\CC^2)^N$ decomposes now into sum of one "bad" 8-dimensional
representation $(\rho_2,\rho_1)$  of type~I and four another 
"good" representations $2\rho_2+2\rho_1$ of type~II~\cite{PS}.
The type~II representations are isomorphic to the ordinary $U(sl(2))$
ones. What about the type~I representation $(\rho_2,\rho_1)$ it can be
considered as a result of gluing of two representations $\rho_2$ 
and $\rho_1$. This  $(\rho_2,\rho_1)$ representation is
indecomposable,
but is not irreducible for it contains 3-dimensional invariant subspace.

In general $q^{p+1}=-1$ case~\cite{LUS,PS} the configuration space
splits into sum of ``bad" type~I representations with highest weights
$S_z\geq p/2$ and ``good" type~II repr-s with highest weights
$S_z< p/2$ which are simultaneously not subspaces of some ``bad" ones.
The highest weight vectors $v_j$ of the good repr-s can be
characterized~\cite{PS} by the following condition 
\begin{equation}
 \label{eq:f}
   v_j \in V_p\equiv \Ker X/\Im X^p   	 
\end{equation}	   
Because of $U_q(sl(2))$ invariance of $H_{xxz}$ we can restrict
its action on the space $V_p$.
We will call the result of this Quantum Group reduction as
the Minimal Model of Integrable Lattice Theory ($LM(p,p+1)$)
because its termodinamic limit is $M(p,p+1)$, the
ordinary Minimal Model of CFT with the Virasoro central charge
$c=1-\frac{6}{p(p+1)}$~\cite{ALC,PS,ZH}.
   
\section*{III. The Quantum Group reduction and the truncation\\
of the fusion functional relations}

The XXZ model(\ref{eq:b}) was solved by Alcaraz et al \cite{ALC}
using the coordinate Bethe ansatz method. Sklyanin constructed the
family of transfer-matrices $T(u)$~\cite{CHER,SKL,MN} commuting 
between themselves and with $H_{xxz}$. It was shown in \cite{KUL}
that  Sklyanin's transfer-matrix $T(u)$ commutes with the Quantum
Group $U_q(sl(2))$. So  the action of $T(u)$ as well as $H_{xxz}$ 
can be restricted on the  space $V_p$ if $q^{p+1}=-1$.
To perform this Quantum Group reduction for $T(u)$ we will use
Baxter's $T$-$Q$ equation~\cite{ZH}
\begin{equation}
 \label{eq:g}
 t(u)Q(u)=\phi(u+\eta/2)Q(u-\eta) + \phi(u-\eta/2)Q(u+\eta)
\end{equation}  
where $\phi(u)=\sin 2u\,\sin^{2N}u$ and $t(u)=\sin 2u\,T(u)$
and $Q(u)$  are  eigenvalues of Baxter's auxilary matrix $\hat Q(u)$
commuting with $\hat T(u)$.

The eigenvalue $Q(u)$ for the eigenvector with $M=N/2-S_z$ reversed spins 
has the following form
\begin{equation}
 \label{eq:h}
 Q(u)= \prod\limits_{m=1}^M \sin (u-u_m) \sin (u+u_m)
\end{equation}  
The equation (\ref{eq:g}) is equivalent to Bethe ansatz
equations~\cite{MN}
\begin{equation}
\label{eq:i}
\left[\frac{\sin (u_k+\eta/2)}{\sin (u_k-\eta/2)}\right]^{2N}=
\prod\limits_{m\not=k}^M \frac{\sin (u_k-u_m+\eta)\sin (u_k+u_m+\eta)}
   {\sin (u_k-u_m-\eta)\sin (u_k+u_m-\eta)}
\end{equation}  
for the model (\ref{eq:b}) provided that $T(u)$ has no poles.

The Baxter's equation (\ref{eq:g}) can be thought as a discrete version of
a second
order differential equation~\cite{BLZ}.
So we can look for its second (linearly independent) solution $P(u)$ with
the same eigenvalue $t(u)$ as in (\ref{eq:g})
\begin{equation}
\label{eq:j}
t(u)P(u)=\phi(u+\eta/2)P(u-\eta) + \phi(u-\eta/2)P(u+\eta)
\end{equation} 

It follows from (\ref{eq:g}) and (\ref{eq:j}) that
\begin{equation}
\label{eq:k} 
\frac{Q(u+\eta)P(u)-P(u+\eta)Q(u)}{\phi(u+\eta/2)}=
\frac{Q(u)P(u-\eta)-P(u)Q(u-\eta)}{\phi(u-\eta/2)}
\end{equation} 

If $\eta/\pi$ is irrational it means  that both parts of (\ref{eq:k})
are equal const and we can choose it be $1$. It means just a normalization 
of $P(u)$. Thus we obtain the "quantum Wronskian" condition\cite{BLZ}
\begin{equation}
\label{eq:l}
  Q(u-\eta/2)P(u+\eta/2)-P(u-\eta/2)Q(u+\eta/2)=\phi(u)
\end{equation} 

If $\eta$ is a rational part of $\pi$ the expresions in (\ref{eq:k})
could be equal a periodic function \\ $f(u)$ such that $f(u+\eta)=f(u)$,
but we  assume that in this case $f(u)$ is also equal to $1$.

Inserting (\ref{eq:l}) to (\ref{eq:g}) or (\ref{eq:j}) we obtain
\begin{equation}
\label{eq:m}
t(u)=Q(u-\eta)P(u+\eta)-P(u-\eta)Q(u+\eta)
\end{equation} 

Let us divide (\ref{eq:l}) by $Q(u+\eta/2)Q(u-\eta/2)$, we get
\begin{equation}
\label{eq:n}
\frac{P(u+\eta/2)}{Q(u+\eta/2)}-\frac{P(u-\eta/2)}{Q(u-\eta/2)}=
\frac{\phi(u)}{Q(u+\eta/2)Q(u-\eta/2)}
\end{equation} 

The r.~h.~s.\ of (\ref{eq:n}) a fraction of two trigonometric
polynomials. So there exist the single way to express it as
the following sum
\begin{eqnarray}
\label{eq:o}
\frac{\phi(u)}{Q(u+\eta/2)Q(u-\eta/2)}
&=&
R(u)+
\frac{A(u+\eta/2)}{Q(u+\eta/2)}-\frac{B(u-\eta/2)}{Q(u-\eta/2)}
\end{eqnarray} 
where $R(u+\pi)=R(u)$ is a trigonometric polynomial whose
degree $degR(u)=2N+2-4M$ and $A(u+\pi)=A(u)$ and $B(u+\pi)=B(u)$
are some trigonometric polynomials whose degree are less than
$degQ(u)=2M$.
Then (\ref{eq:m}) can be rewritten as follows
\begin{eqnarray}
\label{eq:p}
\frac{t(u)}{Q(u+\eta)Q(u-\eta)}
&=&
R(u+\eta/2)+R(u-\eta/2)+\nonumber\\
&&{}+\frac{A(u+\eta)}{Q(u+\eta)}-\frac{B(u)}{Q(u)}-
\frac{A(u)}{Q(u)}-\frac{B(u-\eta)}{Q(u-\eta)}
\end{eqnarray} 
 
The term $\frac{A(u)-B(u)}{Q(u)}$ in the r.~h.~s.\ must vanish for
it has extra poles which are absent in the l.~h.~s.\ of (\ref{eq:p}).
So we obtain that $A(u)=B(u)$ and hence
\begin{equation}
\label{eq:q}
\frac{\phi(u)}{Q(u+\eta/2)Q(u-\eta/2)}=R(u)+
\frac{A(u+\eta/2)}{Q(u+\eta/2)}-\frac{A(u-\eta/2)}{Q(u-\eta/2)}
\end{equation} 

Provided that $F(u)$ is  such function that
\begin{equation}
\label{eq:r}
R(u)=F(u+\eta/2)-F(u-\eta/2)
\end{equation} 
we can rewrite (\ref{eq:n})
\begin{eqnarray}
&&\frac{P(u+\eta/2)}{Q(u+\eta/2)}-\frac{P(u-\eta/2)}{Q(u-\eta)}=\nonumber\\
&&\qquad
=F(u+\eta/2)+\frac{A(u+\eta/2)}{Q(u+\eta/2)}-
F(u-\eta/2)-\frac{A(u-\eta/2)}{Q(u-\eta/2)}
\label{eq:u}
\end{eqnarray} 

It means that we can choose the second solution of Baxter's equation
(\ref{eq:j}) in the following form~\cite{PRST}
\begin{equation}
\label{eq:v}
P(u)=F(u)Q(u)+A(u)
\end{equation} 

The new constructed  solution of Baxter's equation is not a periodic 
function in the general case. It is a periodic function if and only if
there exists a periodic solution of (\ref{eq:r}) $F(u)$.

We delay the discussion of the question when such solution does exist
and get now the consequences of its existence provided  it takes place.

Let us define the following function
\begin{equation}
\label{eq:w}
\textstyle
t_k(u)=Q\left(u-{(k+1)\eta\over2}\right)
P\left(u+{(k+1)\eta\over2}\right)-
P\left(u-{(k+1)\eta\over2}\right)
Q\left(u+{(k+1)\eta\over2}\right)
\end{equation}
where $k$ is any nonnegative integer. 

Comparing with (\ref{eq:l}) and (\ref{eq:m}) we see that $t_0(u)=\phi(u)$
and $t_1(u)=t(u)$.
It is easily to obtain~\cite{PRST} from (\ref{eq:w}) the following
functional equations for $t_k(u)$
\begin{equation}
\label{eq:x}
\begin{array}{c}
t_k\left(u+{\eta\over2}\right)t_k\left(u-{\eta\over2}\right)-
t_{k+1}(u)t_{k-1}(u)=
\phi\left(u+{(k+1)\eta\over2}\right)
\phi\left(u-{(k+1)\eta\over2}\right)\\
\\
t_k(u)t_1\left(u-{(k+1)\eta\over2}\right)=
t_{k+1}\left(u-{\eta\over2}\right)\phi\left(u-{k\eta\over2}\right)+
t_{k-1}\left(u+{\eta\over2}\right)\phi\left(u-{(k+2)\eta\over2}\right)
\end{array}
\end{equation}

These functional relations~\cite{KR} coincide with the relations for
eigenvalues of the fused transfer-matrices of the XXZ model received by Zhou 
in~\cite{ZH} and by Behrend, Pearce, and O'Brien in ~\cite {BPB} for
the ABF models with fixed boundary conditions.

Let us return now to the question about existence of the periodic
solution of (\ref{eq:r}).
Recalling that $\phi(-u)=-\phi(u)$ and the definition (\ref{eq:o})
of $R(u)$  we conclude that $R(u)$ also is an odd trygonometric
polynomial whose degree $degR(u)=2N+2-4M \geq 2$ provided
the number of the reversed spins $M$ is not more than $N/2$.
So such polynomial can be written as follows
\begin{equation}
\label{eq:y}
R(u)=\sum_{m=1}^{2N+2-4M}R_m\sin mu
\end{equation}
where $R_m$ are some coefficients which are defined from (\ref{eq:o}).

If $\eta/\pi$ is irrational the periodic solution of (\ref{eq:r}) does exist
and has the following form
\begin{equation}
\label{eq:z}
F(u)=-\frac12\sum_{m=1}^{2N+2-4M}\frac{R_m}{\sin\frac{m\eta}{2}}\cos mu
\end{equation}
But if  $\eta/\pi$ is rational number the situation is more
sophisticated. Namely we will concentrate on the case $\eta=\frac{\pi}{p+1}$
corresponding to $LM(p,p+1)$ model.
In this case we get from (\ref{eq:z})
\begin{equation}
\label{eq:A}
F(u)=-\frac12\sum_{m=1}^{2N+2-4M}
\frac{R_m}{\sin \frac{m\pi}{2(p+1)}}\cos mu
\end{equation}
So if we try to find the solution in sector when $2N+2-4M\geq 2(p+1)$
or equivalently $S_z\geq p/2$ we see from (\ref{eq:A}) that the periodic
solution does not exist. The reason is just a resonance.

Thus we conclude that a periodic $F(u)$ and the second periodic
solution of Baxter's equation exist if we apply it to subspaces
of the configuration space with $S_z<p/2$ !
We see that this inequality coinsides with the condition for
the Quantum Group reduction from the part II . So we arrive to the
main statement of this section:\\
{\bf The second periodic solution of Baxter equation for $\eta=\pi/(p+1)$
case exists if and only if the configuration space of the model
is undergone Quatum Group reduction}.\\
And for this case we have the additional relations for eigenvalues
of the transfer-matrices $t_k(u)$.
The equation (\ref{eq:w}) takes now the following form
\begin{equation}
\label{eq:B}
t_k(u)=Q\left(\textstyle u-\frac{(k+1)\pi}{2(p+1)}\right)
    P\left(\textstyle u+\frac{(k+1)\pi}{2(p+1)}\right)-
    Q\left(\textstyle u+\frac{(k+1)\pi}{2(p+1)}\right)
    P\left(\textstyle u-\frac{(k+1)\pi}{2(p+1)}\right) 
\end{equation} 
So we see that
\begin{equation}
\label{eq:C}
t_p(u)=0
\end{equation}    
and
\begin{equation}
\label{eq:D}
t_{p-1-k}(u)=-t_k(u+\pi/2)
\end{equation}

So we obtain the second main statement of this section:\\
{\bf  Quantum Group reduction of the model is
equivalent to the truncation of functional relations }(\ref{eq:x}).
\footnote{This statement was also checked numerically by J.~Suzuki~\cite{SUZ}.}

\section*{IV. The solution of the truncated functional
           equatios. Case $\eta=\pi/4$.}
	   
In the case $\eta=\pi/4$ (or $q^4=-1$) the truncated 
fuctional equations take the following form
\begin{equation}
\label{eq:E}
t(u+\pi/8)t(u-\pi/8)=\phi(u+\pi/4)\phi(u-\pi/4)-
\phi(u)\phi(u+\pi/2)   
\end{equation}
After substitution $\phi(u)=\sin2u\,\sin^{2N}u$ to (\ref{eq:E})
we obtain
\begin{equation}
\label{eq:F}
t(u+\pi/8)t(u-\pi/8)=2^{2N}[\sin^{2N+2}2u-\cos^{2N+2}2u]  
\end{equation}
These equations have been considered and solved in ~\cite{BPW} 
as the equations for the transfer matrix of Ising model on the
cylinder.

For $T(u)=\frac{t(u)}{\sin2u}$ we can rewrite
the last equation as

\begin{equation}
\label{eq:G}
2^{2N-1} T(u+\pi/8)T(u-\pi/8)=\prod_{k=1}^{N/2}|\varphi_k(u+\pi/8)|^2
|\varphi_k(u-\pi/8)|^2
\end{equation} 
where we suggest $N$ be even and
\begin{equation}
\label{eq:H}
\varphi_k(u)=\sin(2u+\pi/4)-\omega^k\sin(2u-\pi/4) 
\end{equation}  	   
and $\omega=\exp \frac{\i math\pi}{N+1}$.

Using that $\varphi_k(u+\pi/8)=-\omega^k\varphi_k(u-\pi/8)$
we find $2^{N/2}$ real solutions of (\ref{eq:G})
\begin{equation}
\label{eq:I}
T(u)=T_{n_1,\ldots, n_{N/2}}\equiv2^{\frac{1-2N}{2}}
\prod_{k=1}^{N/2}|\varphi_k\left(u+\frac{\pi n_k}{4}\right)|^2
\end{equation}   
where $n_1\ldots n_{N/2}=0,1$.

Another form the last expression is
\begin{equation}
\label{eq:K}
T_{n_1,\ldots, n_{N/2}}=2^{\frac{1-2N}{2}}
\prod_{k=1}^{N/2}\left[1+(-1)^{n_k}\cos \frac{\pi k}{N+1}\cos 4u\right]
\end{equation}   

We recognize in integers $\{n_k\}$ the occupation numbers
of the fermions of one dimensional Ising model.
It is not suprising because $LM(3,4)$ exactly coincides
with Ising model as was mentioned above~\cite{ALC,HAM,ABF}.

\section*{Acknowledgements}

We are grateful to F.~Alcaraz, V.~Bazhanov, B.~Feigin, S.~Khoroshkin,
M.~Lashkevich, S.~Loktev, S.~Lukyanov,P.~Pearce, Ya.~Pugai, J.~Shiraishi and
J.~Suzuki for usefull discussions.A.~B. is very indebted to M.Lashkevich
for his help in the preparation of the TeX-version of this tekst.
We are thankful to Prof.P.Pearce who attracted our attention to works
~\cite{BPB,BPW}.
We would like to thank T.~Miwa for his kind hospitality in RIMS.
This work is supported in part by RBRF--99--02--16687,96--15--96821,
INTAS--97--1312 (A.~B.), RBRF--98-01-00070, INTAS--96--690 (Yu.~S.).

\end{document}